# An Integrated e-Science Analysis Base for Computational Neuroscience Experiments and Analysis

Kamran Munir, Saad Liaquat Kiani, Khawar Hasham, Richard McClatchey, Andrew Branson, Jetendr Shamdasani and the N4U Consortium

*Centre for Complex Cooperative Systems, University of the West of England, Coldharbour Lane, Bristol BS16 1QY, UK*

**Abstract**

**Purpose** - To provide an integrated Analysis Base in order to facilitate computational neuroscience experiments, following a user-led approach to provide access to the integrated neuroscience data and to enable the analyses demanded by the biomedical research community.

**Design/methodology/approach** - The design and development of the N4U Analysis Base and related Information Services addresses the existing research and practical challenges by offering an integrated medical data analysis environment with the necessary building blocks for neuroscientists to optimally exploit neuroscience workflows, large image datasets and algorithms in order to conduct analyses.

**Findings** - The provision of an integrated e-science environment of computational neuroimaging can enhance the prospects, speed and utility of the data analysis process for neurodegenerative diseases.

**Originality/value** - The N4U Analysis Base enables conducting biomedical data analyses by indexing and interlinking the neuroimaging and clinical study datasets stored on the Grid infrastructure, algorithms and scientific workflow definitions along with their associated provenance information.

**Paper type** - Research paper

**Keywords:** Computational neuroscience; data analysis; e-science; data integration; scientific workflow; information service

## 1. Introduction

Unprecedented growth in the availability and accessibility of imaging data of persons with brain diseases has led to the development of computational infrastructures offering scientists access to the image databases. These image databases are utilised through e-science services such as sophisticated image analysis algorithm workflows (also referred to as pipelines), providing access to powerful computational resources, and visualization and statistical tools. Scientific e-infrastructures have been and are being developed in Europe and North America offering a suite of services for computational neuroscientists and seeking convergence towards a worldwide infrastructure that can constitute the foundations of a global *Virtual Imaging Laboratory*. However, existing infrastructures such as the existing work in the EU FP7 project entitled neuGRID (Redolfi et al., 2009), either address the needs of a restricted user group, e.g. *Alzheimer's* imaging neuroscientists working with a specific dataset and application, or are targeted at a highly specialized community.

*1.1. Research Problem*

A facilitated e-science environment needs to be developed where the broader neuroscience community will find a large array of scientific resources and services. Such an environment needs to span the scientific and global challenges of sophisticated automated image analysis on databases of unprecedented size, thereby enabling further progress in the understanding and cure of neurodegenerative (NDD), white matter (WMD) and psychiatric (PSY) diseases. These diseases, which primarily affect the brain, represent a huge burden and grief to individuals and society. Our continuing work in the EU FP7 project *neuGRIDforUsers* (N4U) (Frisoni et al., 2012) provides an e-science environment by further developing and deploying the neuGRID infrastructure to deliver a *virtual laboratory* which will offer neuroscientists access to a wide range of datasets, algorithm applications, computational resources, services and support. The virtual laboratory, whose architecture is explained in Section 2, is not only being developed for imaging neuroscientists but is also being designed to be adaptable to other user communities.

*1.2. Aims and Objectives*

The N4U (https://neugrid4you.eu) virtual laboratory stores image and clinical study datasets, algorithms and pipeline definitions in a Grid-based infrastructure and provides information and analysis services to its user community. A help desk and a user-facing dashboard interface for interactive use supplements this laboratory. To enable access to the datasets, pipelines and algorithms stored in the infrastructure, the virtual laboratory is aided by a set of information and analysis services. These information and analysis services depend on a computational understanding of the stored datasets, pipelines, algorithms, etc., within a so-called N4U *Analysis Base*. The objective of this article is to present the N4U virtual laboratory infrastructure and discuss the challenges faced in indexing of heterogeneous datasets in the N4U analysis base, elicitation of implicit metadata from such datasets, querying support for users and management of provenance data related to user utilisation and interaction with the information services.

*1.3. Research Challenges*

There are multifaceted challenges in designing a virtual laboratory for neuro-imaging and related scientific analysis. Any form of analysis depends on the availability and quality of the source data. In this context of the N4U project, the source data is being collected from various clinical and research institutions across Europe and in North America. The heterogeneity of the datasets, differences in image and result formats, and granularity of the collected information are major challenges in this aspect. Addressing these challenges requires a data harmonization oriented solution so that the information/analysis services that use this data can use standard/homogenous mechanisms for utilizing this data. Overcoming the data heterogeneity challenge can enable proper indexing of the datasets in a central repository, which can serve as the catalogue to the information and analysis services. However, most of the datasets do not have the associated metadata structures that can ease relational indexing in data stores. This challenge is further compounded by ownership, privacy and security concerns related to the clinical datasets. Therefore each dataset has to be individually reviewed before indexing and appropriate accessibility constraints specified in the metadata structures.

The virtual laboratory will be utilized by the users to carry out analyses of clinical data through algorithmic pipelines. Such interactions will result in generation of new data or datasets. User requirements inform us that the users would like to share not only the results of their analysis within the community but also the specification of the processes through which those results are obtained. Such sharing will enable other users within the community to re-run the analysis in order to verify and validate the results of earlier analyses. This specific requirement can be handled by maintaining provenance of each analysis carried out by the users of the virtual laboratory. The recording, management and sharing of this provenance information on top of heterogeneous indexed datasets that have privacy, security and accessibility concerns is a considerable design and implementation challenge. Within the bounds of the challenges discussed above, there are several similar issues pertaining to the indexing of algorithms and pipelines that will be used in carrying out the analyses in the virtual laboratory. Our design and implementation level solutions to address these challenges are encompassed in the N4U *analysis base* component of the N4U infrastructure.

This paper presents the N4U analysis base within the context of the N4U infrastructure and highlights how this work addresses the listed challenges and paves the way for neuroscientists to access the integrated e-science environment of computational neuroimaging, which can enhance the prospects, speed and utility of the data analysis process for neurodegenerative diseases. The N4U virtual laboratory infrastructure is described in detail in Section 2 and the analysis base pre-requisites are presented in Section 3. The analysis base architectural details are presented in Section 4 and a use-case is described in Section 5 that also highlights how the analysis base is utilized by the N4U information and analysis services. We present related work in the area in Section 6 and Section 7 discusses the accomplishments of targeted N4U research challenges. Finally, we conclude this paper in Section 8 and also discuss our future work.

**2. The N4U Integrated Virtual Laboratory Infrastructure**

In N4U, the development of a virtual laboratory infrastructure was identified as a major requirement in providing a facilitated e-science environment where the broader neuroscience community can find a large array of scientific resources and services. The N4U virtual laboratory, whose architecture is illustrated in Figure 1, offers neuroscientists access to a wide range of datasets, scientific pipelines, algorithm applications, computational resources and services. This virtual laboratory is mainly developed for neuroscientists but designed in a way that should remain adaptable to other user communities. Recently, a significant body of work has been carried out in computational neuroscience research and, in particular for studies of Alzheimer's disease (Mueller et. al., 2005). The main body of work in this area includes for example, neuGRID (Redolfi et al., 2009), Neurolog (Wali et al., 2012), LONI (Dinov et al., 2009), CBRAIN (CBRINE, 2012), and BIRN (Grethe et al., 2005). In these efforts data gathering, management and visualisation has been successfully facilitated, however the constituent data is captured and stored in large distributed databases. Such type of data management, even with the availability of dedicated and powerful software tools and hardware infrastructure, makes it unrealistic for clinical researchers to constantly review, process and then analyse dynamic and potentially huge data repositories for research. In future, it is very likely that such data volumes and their associated complexities will continue to grow, especially due to the increasing digitization of (bio-) medical data. Therefore, users need their data to be more accessible, understandable, usable and shareable. Moreover, the risk at present is that the general lack of integration of the underpinning data infrastructures with user-defined interfacing services - coupled with the issues of unprecedented growth in data volumes and with information heterogeneity - may stifle research and delay the discovery of potential treatments of major and increasingly common diseases.

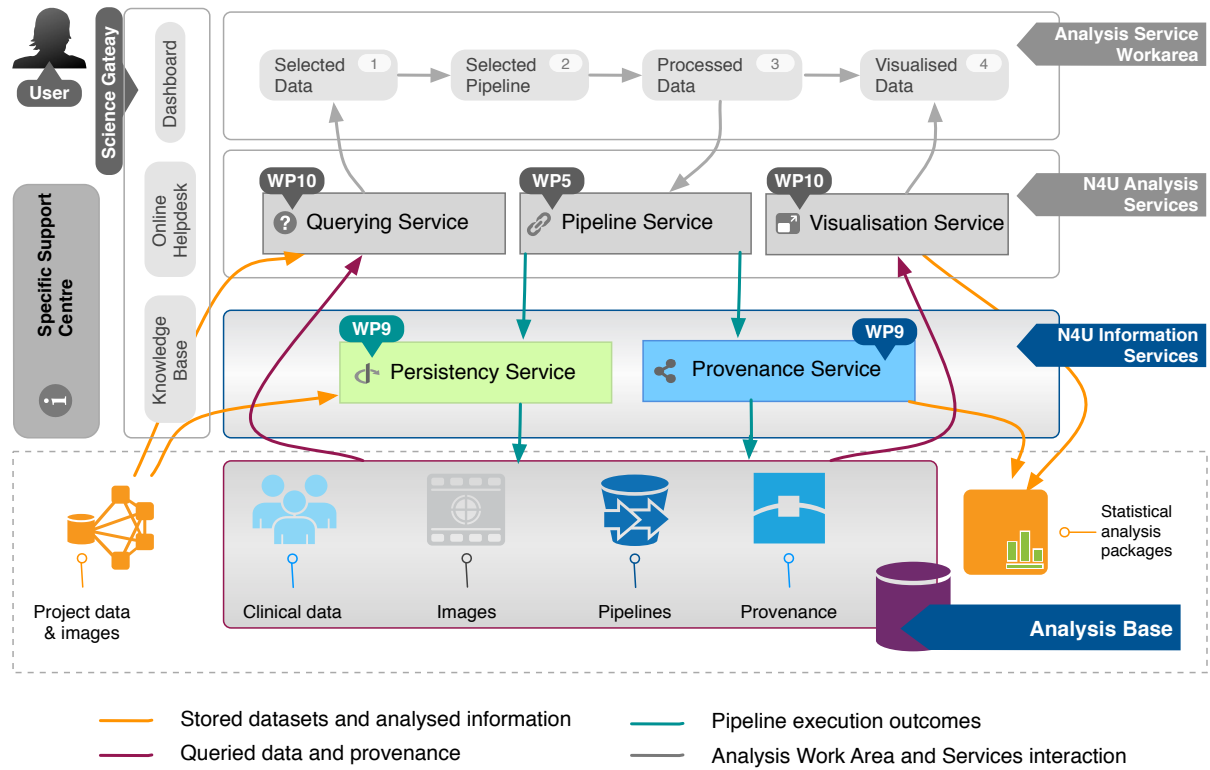

Figure 1: The detailed N4U virtual laboratory built on top of the N4U analysis base

The N4U virtual laboratory has been designed following a user-led requirements analysis approach to provide access to the infrastructure resident data and to enable the analyses demanded by the biomedical research community. This virtual laboratory enables clinical researchers to find clinical data, pipelines, algorithm applications, statistical tools, analysis definitions and detailed interlinked provenance in a user-oriented environment. This has been achieved by basing the N4U virtual laboratory on an integrated analysis database, which has been developed by following the detailed requirements from both neuGRID and N4U projects. This integrated analysis database is entitled the "N4U Analysis Base" and its conceptual design is depicted in Figure 1. The N4U analysis base addresses the aforementioned practical challenges by offering an integrated medical data analysis environment to optimally exploit neuroscience workflows, large image datasets and algorithms to conduct scientific analyses. The high-level flow of data and analysis operations between various components of the virtual laboratory and the analysis base are also highlighted in Figure 1. The N4U analysis base enables such analysis by indexing and interlinking the neuroimaging and clinical study datasets stored on the N4U Grid infrastructure, algorithms and scientific workflow definitions along with their associated provenance information. Moreover, once the neuroscientists conduct their analysis by using this interlinked information, the analysis definitions and resulting data along with the user profiles are also made available in the analysis base for tracking and reusability purposes. Building on the analysis base, the N4U virtual laboratory provides the environment for users to conduct their analyses on sets of images and associated clinical data. In-addition to the analysis base, the N4U virtual laboratory comprises the following major components:

*2.1 Information Services*

The Information Services (see Figure 1) comprise the Persistency and Provenance services within the N4U virtual laboratory. Here, the persistency service complements the analysis base by acting as a wrapper for information storage. The persistency service provides appropriate interfaces for storing the meta-data of datasets e.g. ADNI (Mueller et al., 2005) into the analysis base such that these datasets, which are actually stored in their entirety on the N4U Grid infrastructure (or other similar repositories), become indexed in the analysis base. This index can then be used by other services for allowing users to define their analysis and carry out relevant queries using these datasets. In N4U, the scientific pipelines are submitted to the so-called Pipeline Service for execution on the N4U Grid infrastructure. During and after the execution of pipelines, the pipeline service passes the execution results to the provenance service. Finally, this information is stored in the analysis base for re-use and analysis purposes. One of the major benefits of storing provenance within the analysis base is its use in the reproduction of data when needed by the users. This is particularly important in the context of the healthcare domain, where large and distributed computational resources are used for performing clinical experiments.

*2.2 Analysis Services*

The Analysis Service within the N4U virtual laboratory (see Figure 1) provides access to tracked information (images, pipelines and analysis outcomes) for querying/browsing, visualisation, pipeline authoring and execution. The analysis service is a wizard-based interface allowing researchers to easily combine datasets and algorithms for analyses. Users can choose from items registered in the analysis base that they have been given permission to use, which may be public datasets or their own. Once the execution has been completed, they can choose to share the result with other users, or (re-) run the analysis with different parameters or datasets. The Querying Service provides a mechanism for information retrieval from the N4U analysis base. The querying service facilitates the end-users and mainly the N4U software services, such as the analysis service to submit customised queries to retrieve desired information from the analysis base.

*2.3 Analysis Services Work Area*

The Analysis Service Work Area (see Figure 1) is a facility for users to define new pipelines or configure existing pipelines to be run against selected datasets and dispatch to conduct analysis. The work area is the facility that enables users to prototype their analyses by refining data selections and pipelines, by trying out simple tests and ultimately larger experiments and to visualise the results of those tests and experiments. While the Dashboard (see Section 2.4) is the first point of contact for the user to the underlying N4U services, the work area is the first point of experimentation that empowers the users to use the N4U services and infrastructure.

*2.4 Science Gateway and Specific Support Centre*

The Science Gateway (see Figure 1) in the N4U virtual laboratory provides facilities that include a Dashboard (or user interface), an Online Help Desk and several service interfaces for users to interact with the underlying set of N4U services. The Online Help Desk is a one-stop assistance facility with which the user can interact to learn about the N4U virtual laboratory and how to interact with it via the dashboard. The Knowledge Base brings together administration facilities to provide users with information for the management of the N4U infrastructure. The Specific Support Centre provides users with training materials, e-learning opportunities, active support and other educational materials.

## 3. The N4U Analysis Base Modelling Prerequisites and Design Decisions

The N4U analysis base is required to fully support the different components and services of the N4U virtual laboratory. These requirements can be fulfilled by having the capabilities for (a) indexing all external clinical datasets stored on the Grid infrastructure and elsewhere (b) registering neuroscience pipeline definitions and/or associated algorithms (c) storing provenance and user-derived data resulting from pipeline executions on the Grid (d) providing access to all datasets stored on the Gird infrastructure and (e) storing users' analysis definitions and linking them with the existing pipelines and datasets definitions. In order to meet these requirements, specific data structures and software interfaces to access and manipulate these data structures need to be designed and implemented. While the modelling of the analysis base will be described in the next section, we first provide a discussion on some of the major requirements, their associated dependability and some of the important design decisions that we have taken for the implementation of analysis base.

*3.1. Indexing of External Clinical Datasets*

In the N4U setup, the datasets are stored in the Grid infrastructure. In order to make this data available to the users (e.g. neuro- and citizen scientists) of different N4U services, a mechanism is required that is usable by the software components, which will eventually also be used by human users. Providing direct access to these datasets is not ideal, because a user - who may be preparing to carry out an analysis through the analysis services - may need to select part of the dataset based on certain characteristics. Filtering giga-bytes of data at runtime is a non-trivial task, unless it is methodically indexed beforehand.

*3.2. Register Neuroscience Pipeline Definitions and/or Associated Algorithms*

Neuro- and citizen-scientists will use the N4U virtual laboratory to execute complete pipelines or algorithms over different datasets stored in the Grid infrastructure. Therefore, the N4U virtual laboratory is required to provide a list of pipelines (and algorithms) that are executable on the N4U Grid infrastructure. Moreover, these pipelines may have constraints such as being restricted to certain datasets (or formats) as inputs, algorithms that can be employed within those pipelines, etc. These constraints and other characteristics are specified in pipeline definitions. Therefore, the analysis base is required to index the pipelines already registered (or when they become available) in the N4U infrastructure. Consequently, for completeness, other entities such as algorithms and toolkits related to such pipelines are also indexed in the analysis base.

*3.3. Storing Provenance and User-derived Data Resulting from Pipeline Executions*

Any provenance information generated as a result of some user analysis needs to be stored for cross-analysis, verification and comparison purposes. This provenance information should contain references to the dataset items that were used in the analysis, the algorithms employed, pipeline execution and outcome details, user information/annotation etc. Use-cases such as these require not only the datasets, pipeline and algorithm definitions to be indexed with respect to their contents and properties, but also require the storage of provenance information linked to these entities. Therefore, the analysis base provides a mechanism for storage of the provenance information along with relationships/references of datasets and any data derived from the original datasets as a result of user analysis.

*3.4. Exporting Datasets to the Analysis Base*

In order to index the datasets stored in the N4U infrastructure, the datasets are required to be exported into the analysis base. Exporting the whole datasets into the analysis base is not feasible because of their large size. To meet the goals of the analysis base, it is sufficient to create indexes of the datasets, however, creating these indexes becomes challenging when the metadata associated with the datasets is not available and the disparate formats of the various datasets used in the N4U project. To overcome these challenges, a metadata definition and mapping approach within the persistency service (discussed later in Section 4.2) has been devised to index the datasets in the analysis base.

**4. The N4U Analysis Base Architecture and Implementation**

The N4U analysis base is an information catalogue for the users and a central repository for the Analysis and Querying services to perform search queries for different datasets, algorithms and pipelines in N4U. In this section, we further describe the modelling details of the N4U analysis base in terms of providing support to various N4U virtual laboratory services.

Users who are part of the N4U community are able to index pipelines, algorithms and datasets, which are stored by the persistency service, in the analysis base. Only the users who are marked in the system as active are able to perform these actions. The schema has been designed to facilitate enabling or disabling a user in the analysis base. The analysis base schema expects a user or data provider to provide the meta-data information related to any datasets, which may be data or image files. This information, which is minimally comprised of filename, file location, and type, is stored in the analysis base through the persistency service. To ensure the tracking of a dataset's ownership, an identification of the owner is also stored and mapped against the datasets.

Another objective of the analysis base is to store provenance information related to the pipeline and its analysis, data or image files generated as a result of a pipeline execution. This provenance information aids the N4U user community in reviewing the execution of the pipelines and confirming the outcome produced as a result of their analysis. It also provides a mechanism to the users in debugging the problems with their pipelines or the given input, in case of failures. The analysis base has been designed in a way that it separates the pipeline, algorithm specifications with the pipeline, algorithm execution related information, respectively. Thus this enables rich provenance querying, including queries about the general structure of the pipelines, the activities in the pipelines, the links between different pipeline activities and activity results.

When an analysis is conducted via a pipeline, the analysis base provides a mechanism to record the analysis information and its outcome. The analysis information includes 'which pipeline was executed by whom?', 'at what time?', and 'using which input values?', etc. This information is stored along with the input values and execution output values and serves the Analysis and Querying services in building or searching the provenance information. Moreover, any analysis performed on the N4U datasets is also linked with the provenance of these analyses for tracking and re-usability purposes. In the following sections, we describe the N4U analysis base schema and the interaction of these N4U virtual laboratory persistency and provenance services with the analysis base.

*4.1. N4U Analysis Base Modelling*

In order to map the requirements (discussed earlier in Section 3), the N4U analysis base schema has been designed, a simplified version of which is shown in Figure 2. This schema is composed of five main entities, which include User, Pipeline, Algorithm, Analysis, and DataSet and are briefly described as follows:

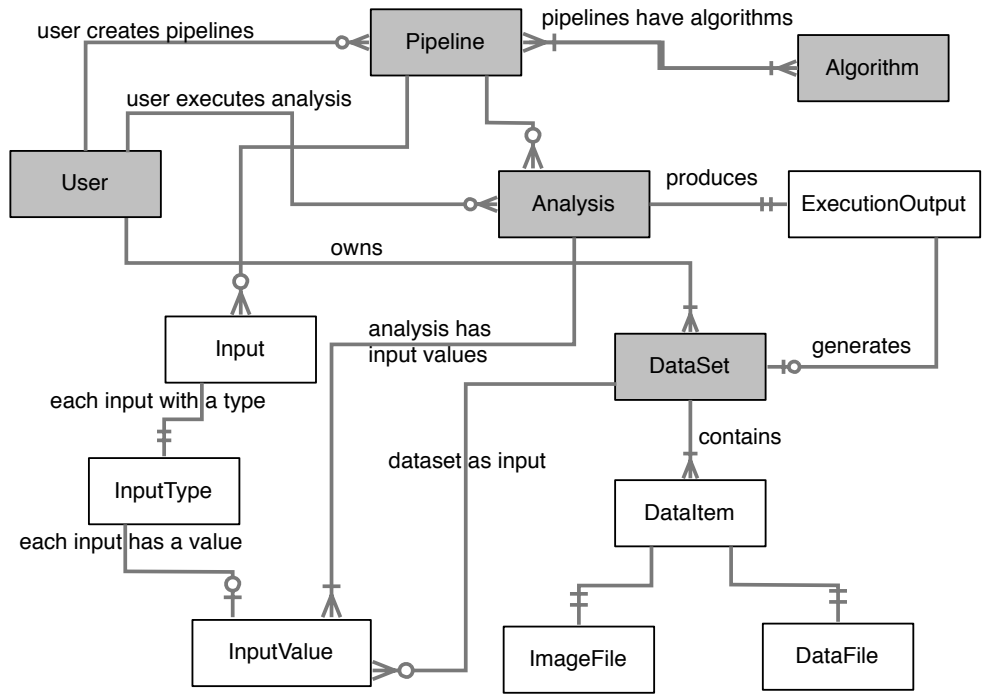

Figure 2: A simplified schema of the N4U analysis base database

- User: In order to provide ownership of activities and data, user information is maintained in the N4U analysis base. A user is a neuroscientist who belongs to a particular research organisation collaborating in a neuroscience experiment and has a particular role in the N4U analysis environment. A user can author multiple pipelines, create datasets and execute these pipelines by providing the required inputs.

- Pipeline: A pipeline (or workflow) is authored by a user in a workflow-authoring tool such as LONI. As discussed earlier in Section 3, pipeline definitions are indexed in the analysis base. These indexes provide the LFN (Logical File Name) for the pipelines available in the Grid infrastructure. By storing only the LFNs instead of the actual pipelines written in XML or other formats, the required storage size is reduced. A user can access a pipeline from the Grid infrastructure using these LFNs. Users can modify or update the pipelines created previously, thus creating several versions of the same pipeline. The N4U analysis base schema also supports this by maintaining version information of each pipeline.

- Algorithm: Each pipeline is composed of different algorithms, which can be thought of as tasks, to perform different types of processing on the neuro-images in a neuroscience analysis. In one pipeline, one or more algorithms can be arranged in an execution order along with their dependencies. Similarly, one algorithm can be used in multiple pipelines.

- Analysis: Execution of a pipeline by providing the required input values (or datasets) is termed conducting an analysis in the N4U terminology. Each analysis and its related information such as input values, ownership and outputs are stored in the N4U analysis base for later querying purposes, such as retrieving pipeline provenance. The values used for each pipeline input are stored in the InputValue table as shown in Figure 2. This information represents the datasets or values for other input types used in a particular analysis. In a data analysis infrastructure that aims to facilitate collaborative research, it is desirable that a user may execute a pipeline authored and shared by another user in order to use or reproduce the analysis results. Therefore, it is pertinent to keep track of the ownership of the pipeline and the analysis since this information is necessary in generating the provenance information of a pipeline, which is essential to verify and reproduce an analysis.

- DataSet: The files used as inputs for analysis or produced as outputs during an analysis are termed a dataset. A dataset can have a single file or a set of files. A set can be composed of files of a single type such as an image file or it can be a combination of files such as image files and data files. The relationship between a dataset and its image or data files is shown in Figure 2. In the N4U virtual laboratory, these datasets are defined and owned by a user; therefore the analysis base schema maintains the relationship between datasets and its user.

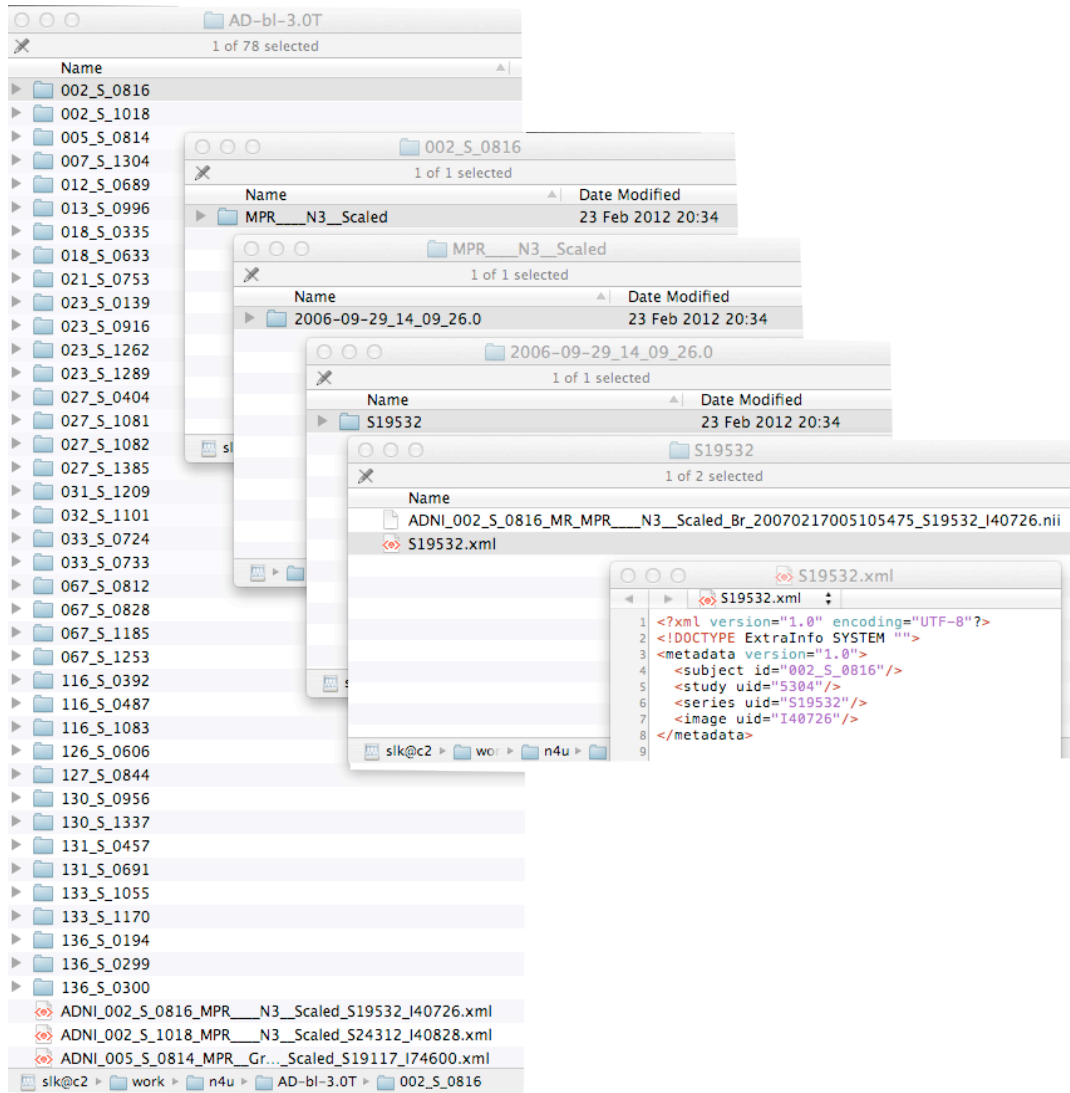

Figure 3: The structure of a data item within a particular study stage in the ADNI dataset

*4.2. Indexing of N4U DataSets, Pipelines and Algorithms through Persistency Service*

The persistency service provides appropriate interfaces for storing the meta-data of datasets (e.g. the ADNI datasets) into the analysis base such that those datasets, which are actually stored in their entirety on the N4U Grid infrastructure, become indexed in the analysis base. This index can then be used by other services for allowing users to define their analysis and carry out relevant queries using these datasets.

Exporting the whole datasets into the analysis base is not feasible because the datasets will be prohibitively large in size. To meet the objectives of the analysis base, only indexes to the datasets need to exist inside the analysis base. Creating these indexes is not straightforward, especially in the absence of any metadata associated with the datasets and the differences in the formats of the various datasets being considered in N4U. Consider the ADNI dataset (Hinrichsa, 2009) - the datasets may be stored in the Grid infrastructure in an archived format, which limits the indexes to simple URL links to the archived files of the dataset. This format of indexing has poor usability because it does not provide a good characteristic-based filtering of the dataset contents. Even if the dataset is in an un-archived format, the indexing process is not simple. The characteristics needed to index the contents of the dataset may be in formats that require pre-processing (e.g. in separate metadata files) or not exist at all.

In the case of ADNI dataset, the contents are organised in multi-level folder structures that contain outputs of different clinical studies at different stages. Each stage (see Error! Reference source not found.) contains dozens of image outputs that are organised in multi-level directory structures along with the associated clinical subject and study data in XML format. There is no formal meta-data associated with the dataset as a whole. To automate the process of indexing such a dataset, and to accommodate any changes to the original data source, it is essential to devise a metadata structure that is uniform across various revisions of the dataset. In the absence of such a metadata format in the ADNI dataset, we have devised one. This metadata format is specified in an XML Schema Definition (XSD), shown in Figure 4, and models the items within a dataset as a (related) collection of images and data files (files containing study and subject information).

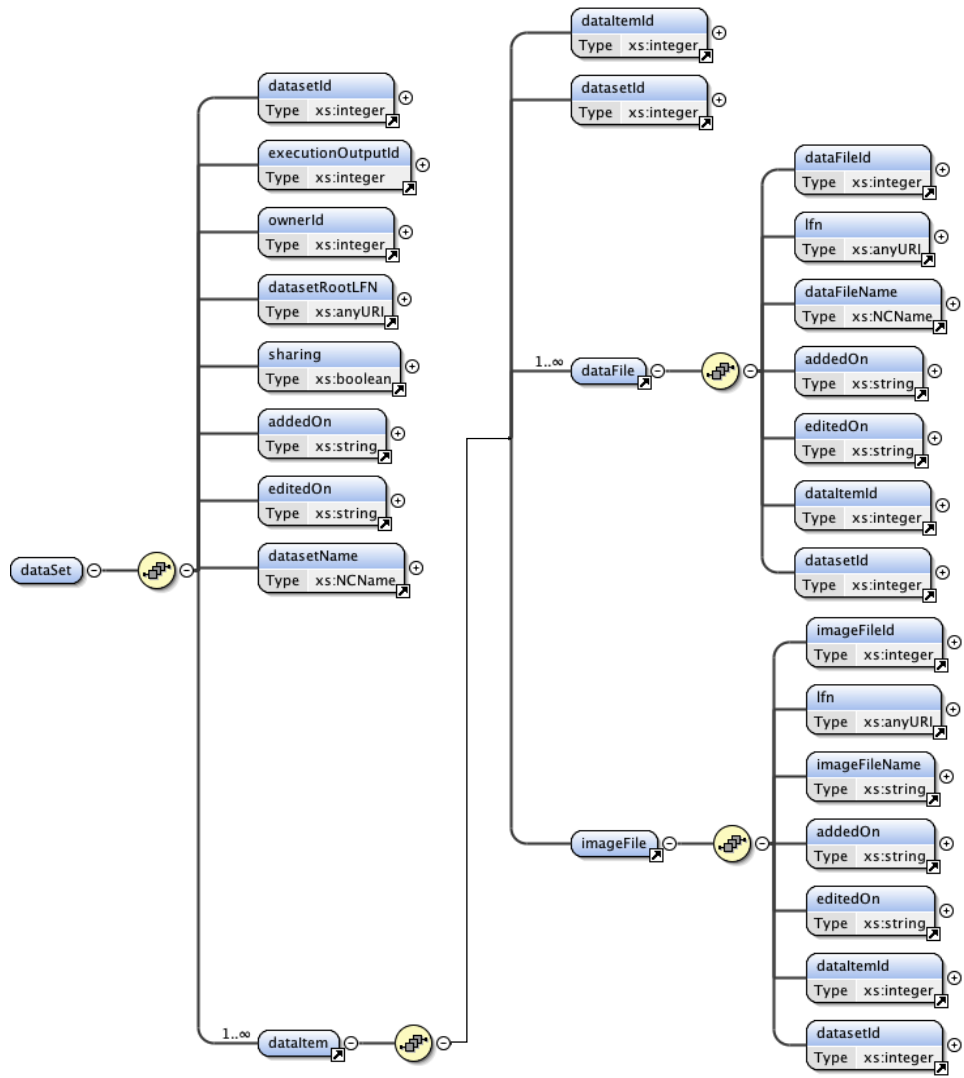

Figure 4: The XSD model for generating the ADNI dataset metadata

The metadata for the ADNI dataset is generated through a software service entitled the DatasetCrawler, which browses (crawls) through the dataset directories (on the Grid infrastructure) and records the structure, names, properties and URLs to the files included in the dataset contents. The DatasetCrawler then generates metadata XML files that conform to the metadata schema devised by the persistency service. These metadata files are exported to the persistency service, which then indexes the metadata in the analysis base. Thus, the persistency service complements the N4U analysis base, acting as a wrapper for storing indexes into the analysis base. The process of generating data metadata is algorithmically described in Algorithm 1.

**Algorithm 1:** Generating metadata from an on-disk dataset

```
WHERE DATASET is dataset containing Image (IMG) and Data (DAT) files
WHERE DS is the data structure that models that DATASET
WHERE DI is the data structure that models the items contained in a DATASET
WHERE DatF is the data structure that models the data files in a DI
WHERE ImgF is the data structure that models the image files in a DI

DS = φ
Switch to the root directory of the DATASET
FOR EACH sub-folder χ in DATASET
    DI = φ
    DatF = φ
    ImgF = φ
    FOR EACH file ω in χ
        IF ω type of IMF
            ImgF = ImgF ∪ ω
        IF ω type of DAT
            DatF = DatF ∪ ω
    DI = DatF ∪ ImgF
    DS = DS ∪ DI
RETURN DS
```

The persistency service also allows the storage of pipeline execution outcomes e.g. one carried out by a user. Similarly, user-annotated data or knowledge-derived by the user through analysis/data annotation are also stored in the analysis base through interfaces provided by the persistency service. The operation of the persistency service and its interaction with the analysis base is shown in Figure 5:

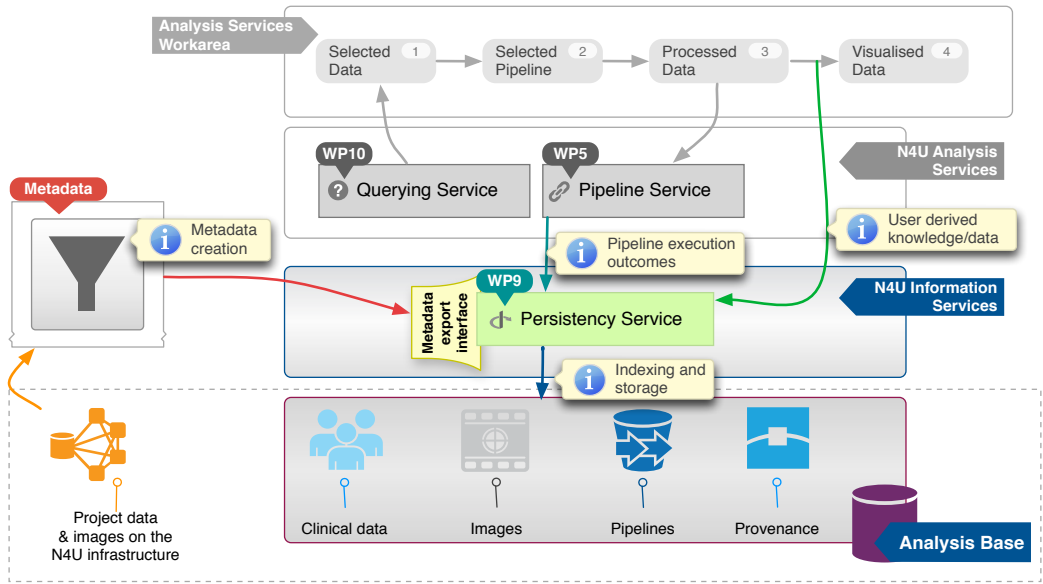

Figure 5: The N4U persistency service and its interaction with the analysis base

*4.3. Provenance Management and Storage Via Persistency Service*

Simply creating and executing pipelines is not sufficient on its own; it is important that results, as and when required, can be reproduced and reconstructed using past information. This leads to the design and development of a dedicated service called the 'provenance service' for this task. The provenance service keeps track of the origins of the data and its evolution between different stages and services. This service allows users to query analysis information, to regenerate analysis workflows, to detect errors in past analyses and to validate the results. Such a service supports and enables the continuous fine-tuning of the pipelines in the N4U project by capturing (1) pipeline specifications; (2) data or inputs supplied to each pipeline component; (3) annotations added to the pipeline; and (4) execution outputs or errors generated during analysis.

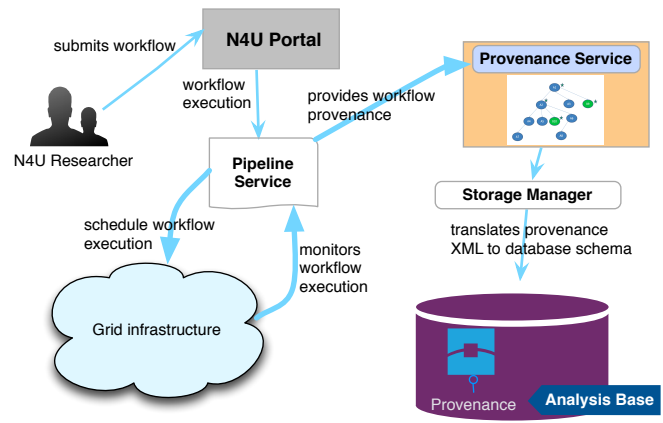

Figure 6: Detail of provenance capturing in the N4U virtual laboratory

Figure 6 shows an overview of provenance capturing in the N4U virtual laboratory. A researcher submits a pipeline through the N4U portal (a graphical user interface) for execution on the Grid infrastructure. The submitted workflow defines the flow and inter-dependencies of activities i.e. algorithms and their inputs. Upon receiving the workflow, it is then passed onto the pipeline service. The pipeline service is responsible for parsing the submitted workflow for consistency checks and for identifying suitable resources on the Grid for its execution. The pipeline service creates a scheduling plan for the workflow and schedules the workflow to suitable resources available on the Grid. This service is also responsible for periodically monitoring the execution status of the submitted workflow. In the case where a workflow fails during execution due to any runtime problem, this service will react and reschedule the workflow to another Grid resource. Along with status monitoring, the pipeline service retrieves the execution output along with other execution related details and update the provenance service. The provenance service exposes interfaces to record provenance information of a pipeline and record it along with the output and error logs in the analysis base via a Storage Manager component. The Storage Manager translates the information coming from the provenance service and maps it to the underlying analysis base schema.

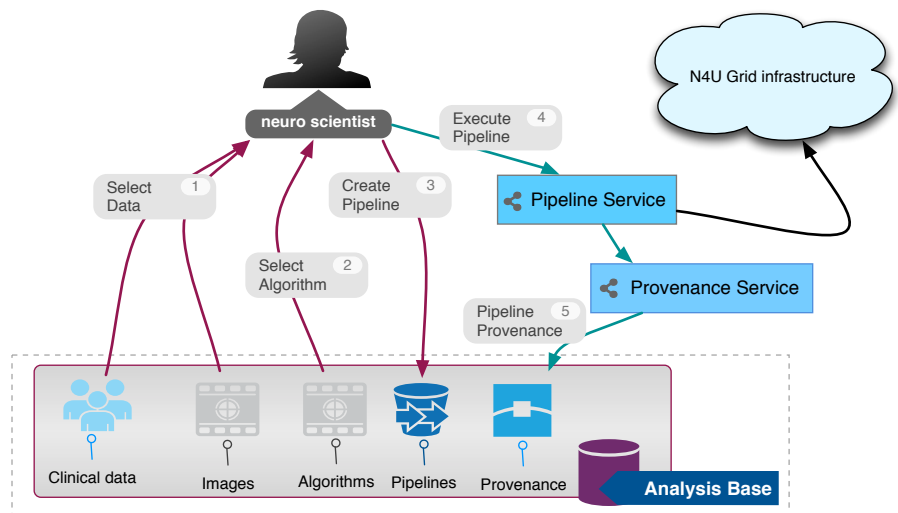

Figure 7: A neuroscientist using the N4U analysis base for his analysis

## 5. Example N4U Neuro-scientists Case Studies using the Analysis Base

Consider, with the help of Figure 7, a use-case that briefly elaborates the role of the analysis base during a user's interaction with the N4U virtual laboratory. A neuroscientist can exploit the capabilities provided by the N4U infrastructure to explore the available neuro-images and their associated clinical study data, and then choose a subset according to her selection criteria. There may be thousands of images indexed in the N4U analysis base, however, she may be interested in a few images that conform to her analysis criteria, i.e. neuro-images of the male patients with age above 50 years and clinical history of at least 2 assessments. Using the rich querying mechanism provided by the N4U analysis base, she is able to find a set of images conforming to her search criteria that she can use in her neuro-analysis. Once the images are chosen, she selects the algorithms to process and analyse them. According to the nature of her analysis, she arranges the algorithms in a sequence and defines their inputs, a process known as pipeline authoring, to achieve the desired result from the analysis. Once a pipeline has been authored, a reference to the pipeline is stored in the N4U analysis base and it is then submitted to the N4U Grid infrastructure for execution through the pipeline service (as shown in Figure 7). An entry of this analysis is recorded in the N4U analysis base. The execution of analysis will produce logs and outputs, which are collected as part of the analysis information for user review. Besides this, the execution times and behaviour of the analysis process are also stored in the N4U analysis base for providing pipeline provenance information to the user. This information is collected using the provenance service. With the provenance information, a user can debug his analysis, verify the output and measure the execution performance of his analysis. These steps and user interaction with the N4U analysis base are illustrated in Figure 7.

## 6. Related Work

There has been significant research carried out in exploiting the Grid infrastructure and data management approaches to store and analyse the neuroimaging data for the neuroscience community. The NeuroLOG (Franck et al., 2010) project proposes a federated approach to integrate neuro-image files, associated metadata and neuroscience semantic data distributed across different sites. A Data Management Layer (DML) has been devised to hide the underline complexity of data stored in different formats at different sites. In doing so, an additional NeuroLOG database, structured according to defined ontologies, is deployed on each site to achieve complete autonomy of the sites and to facilitate existing data formats stored on the sites. Their main focus is on the storage and semantic retrieval of the data stored across different sites, however this approach does not provide provenance information of the neuroimaging analyses conducted over the neuro-images.

FBIRN (Burak et al., 2010) presents an extensible data management system for clinical neuroimaging studies. They have developed a distributed network infrastructure to support the creation of a federated database consisting of large sample neuroimaging datasets. FBIRN also incorporates the provenance information of an analysis. However, this provenance information does not provide an insight about the execution times of an analysis. Furthermore, the federated multi-site data organization approach requires the configuration and deployment of the proposed framework at each site, which increases the overall complexity. In comparison our proposed approach is comparatively simpler since it provides interfaces for external data sources that can be used to index external data according to a single schema, thus providing a uniform storage view to the users and external data sources. All the transformation and complexity is hidden from external data sources and handled internally by the persistency service (see Section 4.2). In doing so, a data format has been devised for the data providers in order to import their data in the N4U analysis base.

The CNARI framework (Small et al. 2009), used at the Human Neuroscience Laboratory at the University of Chicago, presents a database driven architecture that combines databases for storing the fMRI data and workflows for analysis purposes. It proposes the use of relational databases instead of traditional file-based approach for storing and querying fMRI data (Hassan et al. 2008). It creates separate databases for each experiment and several tables to store images or user-specific data, which require a huge storage space. In order to execute the workflows, CNARI employs SWIFT as a workflow engine (Zhao et al. 2007) and exploits its provenance tracking capabilities to maintain provenance information for reproducing an analysis. Unlike CNARI, our proposed schema stores references to the files available on the N4U Grid infrastructure, thus making it storage efficient. Moreover, it provides runtime provenance information of each analysis, which is a task under progress in SWIFT.

The Extensible Neuroimaging Archive Toolkit (XNAT) (Marcus et al., 2007), developed by the Neuroinformatics Research Group at Washington University at St. Louis, is aimed at offering researchers an integrated environment for archival, search and sharing of neuroimaging datasets. It relies on an extensible XML schema to represent imaging and experimental data and supports a relational database backend. It is aimed at managing large amounts of data via a three-tier design infrastructure consisting of a client front end, the XNAT middleware and a data store. The data store is composed of a relational database and a file system on which images are stored. Our proposed schema is similar to XNAT by storing pointers to the files in the database. However, XNAT only focuses on storing the data whereas our approach not only supports the indexing of datasets but also indexes pipelines and their analyses, and stores the associated provenance information.

GridPACS (Hastings et al., 2005) supports distributed storage, retrieval and querying of image data and associated descriptive metadata. Workflows and metadata are modelled as XML schemas unlike the relational database schema approach used in N4U analysis base, which supports rich querying using SQL. It integrates image data and metadata to maintain provenance information about how the imagery has been acquired and processed. SenseLab (SenseLab, 2012), developed at Yale University, is a metadata driven system to store scientific data using an entity-attribute-value with classes and relationships representation in a relational database. Most of these approaches primarily focus on the storage of clinical data, however, the N4U analysis base not only provides a schema to store and retrieve image datasets but also contains a layout for storing pipelines, analysis and the processed data such as provenance.

## 7. Discussion

This section briefly discusses the accomplishment of the targeted N4U research challenges by the provision of an integrated e-science analysis base.

### 7.1. Achieving the challenge of indexing support via Persistency Service

One of the primary challenges in the design and development of the analysis base is to provide an index of neuroimaging datasets that are stored on the Grid infrastructure. We have described the persistency service (Section 4.2) that encompasses the functionality to address this challenge. A generic dataset metadata schema has been designed to capture metadata information from the datasets stored on the Grid infrastructure. The on-disk datasets are analysed through a DatasetCrawler service that monitors the Grid storage for the addition of new datasets or modifications in existing datasets. The DatasetCrawler service generates a metadata file for each dataset that conforms to the metadata schema. This metadata is then exported to the persistency service, which carries out the indexing of datasets by utilizing the metadata information.

Through the mechanism described above, we have addressed the challenge of indexing the datasets in the analysis base, however we foresee a continuing evolution of the practical mechanism involved in the solution. This continuous evolution is required to accommodate new datasets that will be added to the N4U infrastructure in near and long-term future, which may differ in formats that necessitate a change in the metadata schema model.

### 7.2. Achieving the challenge of Provenance Management

Another major aim in the design and development of the N4U analysis base is to keep track of the origins of datasets, their evolution through different stages and exchange between various services. In this regard, we have described the provenance service that provides a mechanism for the storage of such provenance information within the analysis base along with relationships/references of datasets and any data derived from the original datasets as a result of user analysis. The analysis base schema maintains sufficient information to provide answers to provenance queries such as *who authored and executed a workflow?, at what time?, what output was produced by the workflow?, what inputs were used in an analysis to produce a certain output?, whether a workflow authored by a user was executed correctly and what output datasets it has produced?* etc. The querying service that has been built on top of the analysis base allows users to query such provenance information. The resultant data can then be used to regenerate the analysis workflows, to detect errors and to view execution outcomes in past analyses and also to validate them. Thus, by capturing provenance in the analysis base, linking it with original datasets and respective analysis we support and enable the continuous refinement of users' analysis in the N4U virtual laboratory.

### 7.3. Achieving the Analysis Requirements

Another aim of building the N4U analysis base is to enable the neuroscientists to perform detailed examination of the existing or generated medical data (including images, workflows and analysis outcomes) by providing a data access/sharing facility. The analysis base enables neuroscientists to choose from the datasets registered in the analysis base that they have been given permission to use in their analysis. Once the analysis is performed and results are transferred back to the user, the analysis base provides a mechanism to store an analysis and its outcome. Moreover, the users can choose to share the result with other users, or (re-) run the analysis with different parameters or datasets. In this way, we addressed the challenge of archiving the full cycle of users' analysis, starting from the indexing of datasets, workflow and algorithms to analyses specifications and their execution outcomes. This facility has enabled neuroscientists to conduct peer reviews of scientific work and analyses in order to reuse or verify each other's scientific findings.

## 8. Conclusions and Future Work

We have presented the design and development of the N4U analysis base and related information services, which address existing research and practical challenges by offering an integrated medical data analysis environment with the necessary building blocks for neuroscientists to conduct analyses. The N4U analysis base enables such analyses by indexing and interlinking the neuroimaging and clinical study datasets stored on the N4U Grid infrastructure, algorithms and scientific workflow definitions along with their associated provenance information. We have presented a persistency service and its generic approach that mitigates the challenges involved in indexing various types of datasets coming from various data sources. Moreover, once the neuroscientists have conducted their analyses by using this interlinked information, the analysis definitions and resulting data, along with the user profiles, are also made available to the registered scientific community for tracking and reusability purposes. In doing so, we have presented a provenance service responsible for keeping track of the analysis history, origin of the data being consumed or produced and the analysis outcomes. This work has paved the way for neuroscientists to access the integrated e-science environment of computational neuroimaging, which has enhanced the prospects, speed and utility of the data analysis process for neurodegenerative diseases. Our future efforts are geared towards indexing a larger number of clinical datasets and providing interactive visualisation interfaces for the users of the N4U virtual laboratory.

**Acknowledgements**

This work is funded by the EU 7$^{th}$ Framework Programme under the N4U project (reference 283562). The authors take this opportunity to acknowledge the contribution of all the partner institutions in the N4U project to this paper.